\documentclass[12pt]{iopart}
\usepackage{epsfig,harvard,bm}
\usepackage{amsbsy,amssymb}
\usepackage{pslatex}

\newcommand{\isum}%
{\mathop{\hbox{$\displaystyle\sum\kern-13.2pt\int\kern1.5pt$}}}
\renewcommand{\r}{{\bm r}}

  \newcommand{\A}{{\bm A}}

\newcommand{\bt}{\begin{tabular}}
\newcommand{\et}{\end{tabular}}

\newcommand{\er}{\end{eqnarray*}}
\newcommand{\ba}{\begin{eqnarray}}
\newcommand{\ea}{\end{eqnarray}}
\newcommand{\be}{\begin{equation}}
\newcommand{\ee}{\end{equation}}

\newcommand{\bp}{\begin{minipage}}
\newcommand{\ep}{\end{minipage}}

\begin{document}
\bibliographystyle{jphysicsB}

\title {Entropy-based view of the strong field ionization process.}

\author{I. A. Ivanov$^{1,3}$, Chang Hee Nam$^{1,2}$, and Kyung Taec Kim$^{1,2}$}

\address{$^{1}$Center for Relativistic Laser Science, Institute for
Basic Science, Gwangju 61005, Korea}

\address{$^{2}$Department of Physics and Photon Science, GIST, Gwangju 61005, Korea}

\address{$^{3}$Research School of Physics and Engineering,
The Australian National University,
Canberra ACT 0200, Australia}

\ead{igorivanov@ibs.re.kr}

\begin{abstract}

We apply information theoretic entropies of coordinate and velocity distributions in quantum mechanics
for the description of the strong field ionization process. 
The approach is based on the properties of the entropies used in the information theory, viz., their ability 
to gauge the "distance" between the probability distributions and thus to be sensitive 
to the distributions variations. Study of the entropies as functions of time allows thus to visualize conveniently 
evolution of the wave-function of the system undergoing strong field ionization,  and to pin down, in particular, 
the times when the wave-function begins to change appreciably.

\end{abstract}

\maketitle

\section{Introduction}

Tunneling ionization is a process occurring when atom is exposed to 
a strong laser field with atomic and field parameters satisfying the condition 
$\gamma=\omega\sqrt{2I_p}/E_0 \lesssim 1$ \cite{Keldysh64,kri} 
(here $\omega$, $E_0$ and $I_p$ are the frequency, field strength and ionization potential
of the target system expressed in atomic units). The models introducing notion of the electron trajectory 
proved to be of great utility in explaining many features of this process.
The well-known simple man model (SMM)\cite{hhgd,Co94,KK93,kri,tipis,arbm}, for instance, 
reproduces many qualitative features of the strong field phenomena, such 
as high harmonic generation, attosecond pulse generation and above threshold ionization. 

Such models are capable to render a quantitative description of the 
tunneling ionization phenomena as well. The semi-classical 
TIPIS  tunnel ionization in parabolic coordinates with induced dipole and Stark shift)  model \cite{tipis,cusp3,cmtc1}, for example,
has been shown to produce
accurate quantitative results for the electron spectra \cite{tipis,cusp3,arbm,tipis_naft,landsman2015}.
In this approach, the quantum-mechanical Keldysh theory \cite{Keldysh64} and its modifications \cite{Keldysh64,Faisal73,Reiss80,ppt} 
are used to set up initial velocities for the classical electron motion \cite{cusp3,arbm,tipis_naft}.
Initial value of the coordinate is defined either by the simple Field Direction Model (FDM) or the more 
refined approach using the parabolic coordinate system \cite{landsman2015,bhu}.

Yet the question of what physical reality corresponds to this undoubtedly extremely fruitful
semi-classical picture, or more precisely how we can better visualize  development of the 
ionization process  in time, remains, to some extent, open. 
The notion of the tunneling electron escaping from
under the barrier at a given time, at some point in space, and with certain velocity, is difficult to 
reconcile with conventional quantum mechanics (QM).
One might try to bypass the problem by escaping the conventional QM framework, 
using, for example Bohmian approach \cite{landsman_bohm,bohm_us}, which reintroduces notion of the 
classical trajectory in QM. This solution may be not entirely satisfactory, the Bohmian interpretation being not
universally accepted. 

The so-called quantum orbits introduced by the Imaginary Time
Method (ITM) \cite{ppt,tunr2} seem to provide a basis for such a reconciliation in the framework of the conventional QM.
Motion of the electron in the under-the-barrier region 
in the ITM approach can be visualized as a motion in the
complex time with the complex velocity.  Electron trajectory 
in the under-the-barrier region starts at a 
saddle-point at a complex moment of time  and descends 
on the real axis. At the time when the trajectory crosses the
real time-axis  electron's coordinate and velocity (for the 
most probable trajectory) become real. 
One might then interpret this instant of time 
 as the time of the electron's exit 
from under the tunneling barrier.
This interpretation, however, is not entirely flawless. Because of the analyticity  of the electron's action
as a function of  time 
the path connecting the saddle point and a final moment of time when electron's velocity is
measured, is not unique \cite{tunr2} and must be adequately 
chosen y to take into account the topology of the branch cuts 
arising  when atomic potential is included in the theory
\cite{cpath1,cpath2,cpath3}.  



More in line with the framework of the conventional QM is the point of view that
all information about the system we may obtain or we may need is contained in the wave-function, the question 
is only how to extract it. For instance, one interpretation of the key notion in the theory of 
tunneling ionization- the so-called Keldysh tunneling time, is 
the time necessary for establishing
of the static-field ionization rate when the electric field turns
on instantly \cite{time_adjust}. It characterizes, therefore, the time the wave-function needs to adjust itself.

To describe an adjustment of the wave-function when electric field is present in quantitative terms 
we need a convenient characteristic which
can be used to follow the changes the wave-function undergoes during the ionization process. 
We describe below the
use of the information theoretic Shannon entropy \cite{entropy_shan} 
as such a characteristic. The meaning of the information entropy we need here is different from the 
meaning of the von Neumann entropy in quantum mechanics, which
provides a measure of the purity of the wave function, and which can be used to characterize 
entanglement in the strong field ionization process \cite{entropy7}.
The property of the information entropy we use is its ability to supply
the space of the probability distributions with a metric \cite{entropy_leip,entropy_metr}. In other words,
we use information entropy to quantify the statement that one probability distribution is different from another,
or, which is actually the goal of the present paper, to visualize the development of the strong field ionization process in time,
using entropies as convenient measures illustrating  evolution of the wave-function. 
We will study below entropies of the coordinate and velocity probability distributions for the process
of strong field ionization.   Entropy of a probability distribution $f$ is its integral characteristic assigning a number $S(f)$ - the entropy of the distribution,  to the given distribution.  Evolution of the entropies in time is, therefore, easier to follow, visualize and interpret than evolution of the  wave function. 
The correspondence $f \to S(f)$,  as we shall see,  preserves many important features of the 
ionization process, and allows to illustrate them in a clear and concise manner.

\section{Theory}

Using the definition of the theoretic Shannon entropy \cite{entropy_shan}, 
entropies of the coordinate and velocity distributions
($S_x(t)$, and $S_v(t)$, respectively), can be defined as:

\begin{eqnarray}
S_x(t)= -\int  |\Psi(\r,t)|^2 \log{|\Psi(\r,t)|^2}\ d\r \ ,\nonumber \\ 
S_v(t)= -\int  |\tilde\Psi({\bm q},t)|^2 \log{|\tilde\Psi({\bm q},t)|^2}\ d{\bm q}  .
\label{ent}
\end{eqnarray}

Here $\Psi(\r,t)$ is the coordinate wave-function describing the system, $\tilde \Psi({\bm q},t)$ its Fourier transform. 
Both $\Psi(\r,t)$ and $\tilde \Psi({\bm q},t)$ are not 
dimensionless quantities so, strictly speaking, logarithms of
these quantities in \Eref{ent} are not well defined. 
 For these logarithms to make sense we could 
 choose a parameter  $c_x$ having the physical dimension of 
$|\Psi(\r,t)|^2$, and a parameter  $c_v$ having the physical dimension of 
$|\tilde\Psi({\bm q},t)|^2 $, and 
replace the logarithms in \eref{ent} with 
$\log{(|\Psi(\r,t)|^2}/c_x)$ and 
$\log{(|\tilde\Psi({\bm q},t)|^2/c_v)}$.  The entropies 
would then become functions 
$S_x(t,c_x)$, $S_v(t,c_v)$
of these parameters .  One can easily see that 
since both coordinate and momentum wave-functions are normalized to
unity and remain normalized in the course of the evolution, 
for the entropies obtained using different sets of 
parameters $c^{(1)}$ and $c^{(2)}$ one
has $S(t,c^{(2)})- S(t,c^{(1)})= \log{(c^{(2)}/c^{(1)})}$.  The 
coordinate and velocity entropies
are, therefore, defined by the \Eref{ent}  only up to an 
arbitrary additive factors (the property they share with 
the statistical entropy in the classical physics \cite{LL5}).  
If, as we will do below, we are interested in the entropy
change, this arbitrary factors become immaterial, and 
we can just choose them to be 1 both in the coordinate and the momentum
spaces.

We note  that the definitions of $S_x$ and $S_v$ in \Eref{ent} apply in both length (L)  and velocity (V) gauges we may use to describe atom-field
interaction \cite{Sobelman72}. 
Obviously, $S_x$ does not depend
on the choice of the gauge since  transformation (we assume dipole approximation) 
$\Psi_V(\r,t)=e^{-i\A(t)\cdot\r} \Psi_L(\r,t)$
from the $L-$ to the $V-$ gauge does not affect $|\Psi(\r,t)|^2$. In the case of the 
velocity distribution, velocity ${\bm v}$ in the $V-$ gauge is related to the wave-vector ${\bm q}$ of the Fourier transform $\tilde \Psi({\bm q},t)$
as ${\bm v}= {\bm q} + \A(t)$ . Therefore, velocity distribution in the $V-$ gauge is given by $|\tilde \Psi({\bm v-\A(t)},t)|^2$, 
and the integral defining $S_v(t)$ can still be written in the form \Eref{ent} by shifting the dummy 
integration variable. 

Below we will study the  evolution of $S_x(t)$ and $S_v(t)$ with time for the 
process of strong field  ionization of hydrogen
atom.  The coordinate wave-function $\Psi(\r,t)$ describing the ionization process is obtained 
by solving the time-dependent Schro\"dinger equation (TDSE) for a hydrogen atom in  presence of a laser pulse:

\begin{equation}
i {\partial \Psi(\r) \over \partial t}=
\left(\hat H_{\rm atom} + \hat H_{\rm int}(t)\right)
\Psi(\r) \ ,
\label{tdse}
\end{equation}

with
 $\displaystyle \hat H_{\rm atom} = {\hat{\bm p}^2\over 2}-{1\over r}$,
 and the length gauge form 
 $\hat H_{\rm int}(t)= {\bm E(t)}\cdot \hat{\r} $ 
for the interaction operator. 
We will follow evolution of 
the hydrogen atom initially in the ground state with
the wave-function $\phi_0$ and energy $\epsilon_0=-0.5$ a.u.
To solve the TDSE numerically we follow the procedure
described in detail in  our earlier works \cite{cuspm,circ6}.
We will present below, therefore,  only the most essential details of the calculation.

The electric field of the laser pulse is linearly polarized (along the $z-$ axis, which we use also as the quantization axis) 
and is defined in terms of the vector  potential 
$\displaystyle {\bm E}(t)=-{\partial {\bm A}(t)\over \partial t}$,
where:

\begin{equation}
{\bm A(t)}= -\hat {\bm z} {E_0\over \omega}\sin^2{\left\{\pi t\over T_1\right\}}\sin{\omega t} \ ,
\label{ef}
\end{equation}

with peak field strength  $E_0$, carrier frequency $\omega$, and total duration $T_1=NT$, where 
$T=2\pi/\omega$ is an optical cycle (o.c.) corresponding to the frequency $\omega$ and $N$ is an integer. 
For a given pulse duration $T_1=NT$ we consider evolution of the system on the interval $(0,(N+1)T)$,
allowing one optical cycle of the field-free evolution of the system after the end of the pulse. We will consider pulses 
with the fixed  pulse duration of $N=2$, and
we will vary the carrier frequency $\omega$ and peak field strength $E_0$. The corresponding pulse shapes are shown in \Fref{fig1}.

The solution of the TDSE is represented as a  series in spherical harmonics:

\be
\Psi({\bm r},t)=
\sum\limits_{l=0}^{l_{\rm max}} 
{f_{l}(r,t)\over r}  Y_{l0}(\hat{\r}) \ .
\label{basis}
\ee

Radial variable is discretized  on the  grid with the step-size
$\delta r=0.1$ a.u. in a box of the size $R_{\rm max}$. We used
$R_{\rm max}=600$ a.u. in the calculations below. As for the parameter 
$l_{\rm max}$ in \Eref{basis},  its optimal value depends on the base frequency of the laser pulse,
growing with decreasing frequency. 
Necessary accuracy checks have been performed to ensure that
convergence of the expansion \eref{basis} has been achieved in the calculations.
For the smallest frequency $\omega=0.03$ a.u. we consider below we
had to use $l_{\rm max}=70$.

\section{Results}

\begin{figure}[h]
\begin{tabular}{l}
\resizebox{100mm}{!}{\epsffile{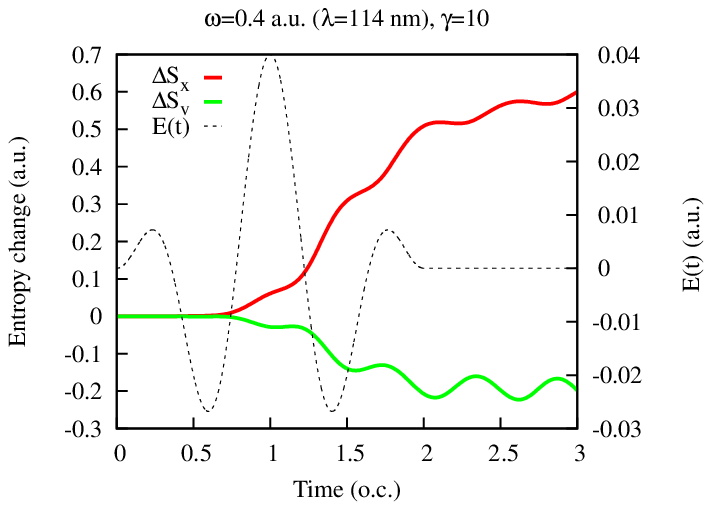}}  \\
\resizebox{100mm}{!}{\epsffile{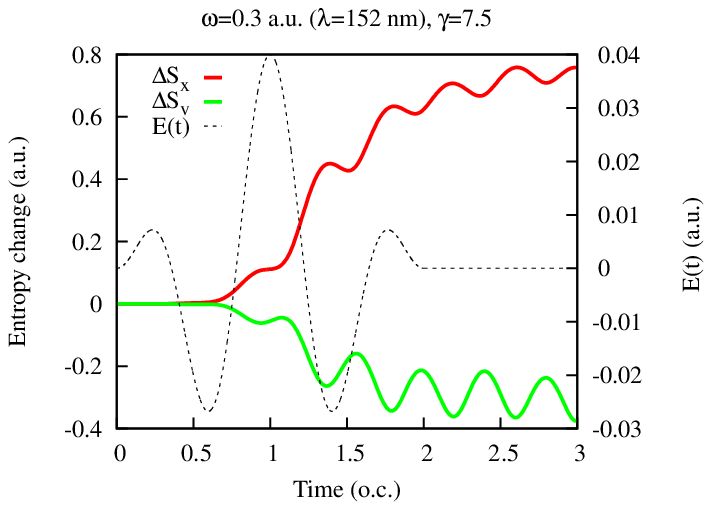}}  \\
\resizebox{100mm}{!}{\epsffile{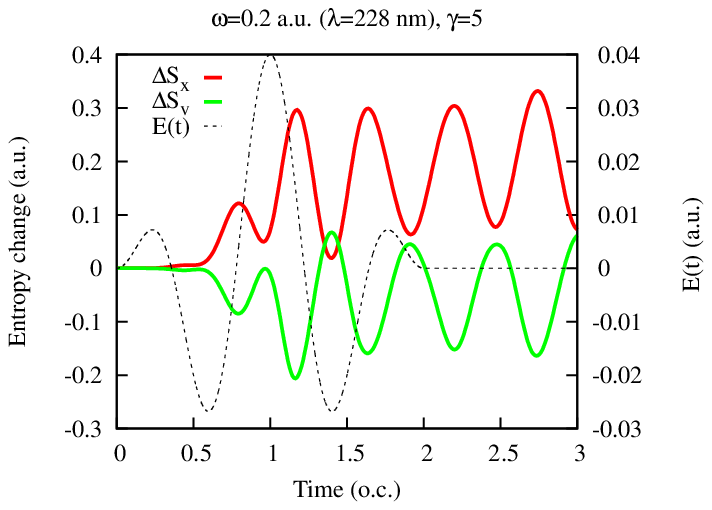}} \\
\end{tabular}
\caption{Color online. Entropies of the coordinate and velocity distributions for a driving pulse
with peak field strength $E_0=0.04$ a.u.  in the multiphoton ionization 
regime.} 
\label{fig1}
\end{figure}

\begin{figure}[h]
\begin{tabular}{l}
\resizebox{100mm}{!}{\epsffile{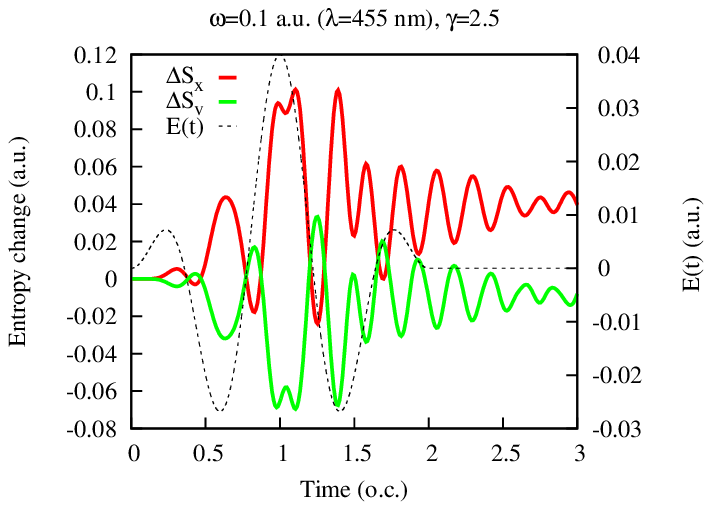}}  \\
\resizebox{100mm}{!}{\epsffile{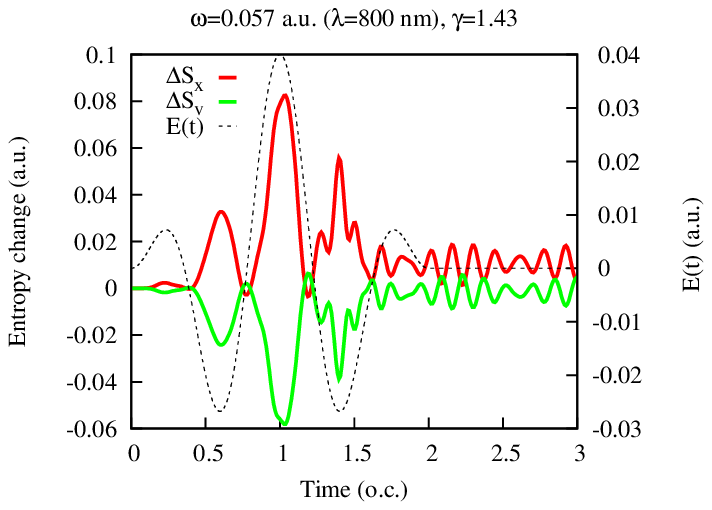}}  \\
\resizebox{100mm}{!}{\epsffile{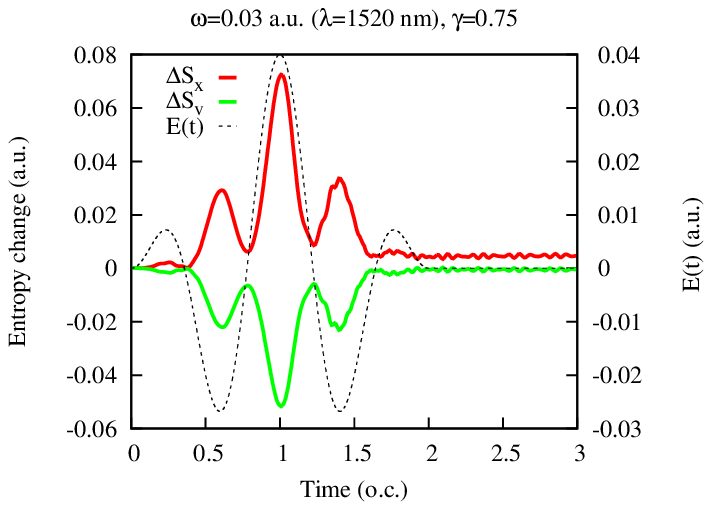}} \\
\end{tabular}
\caption{Color online. Entropies of the coordinate and velocity distributions for a driving pulse
with peak field strength $E_0=0.04$ a.u. in the tunneling 
ionization regime.} 
\label{fig11}
\end{figure}

\begin{figure}[h]
\begin{tabular}{l}
\resizebox{100mm}{!}{\epsffile{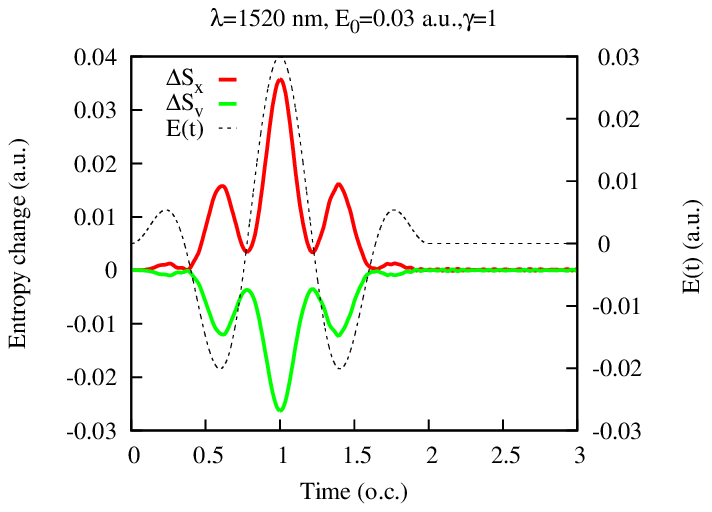}}  \\
\resizebox{100mm}{!}{\epsffile{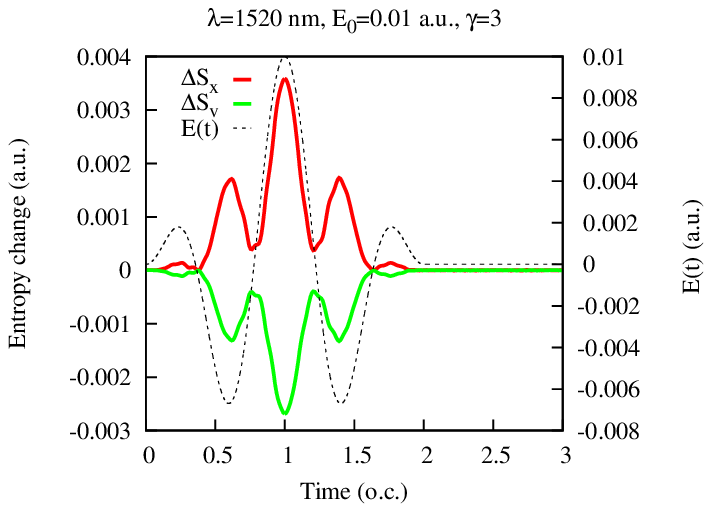}} \\
\resizebox{100mm}{!}{\epsffile{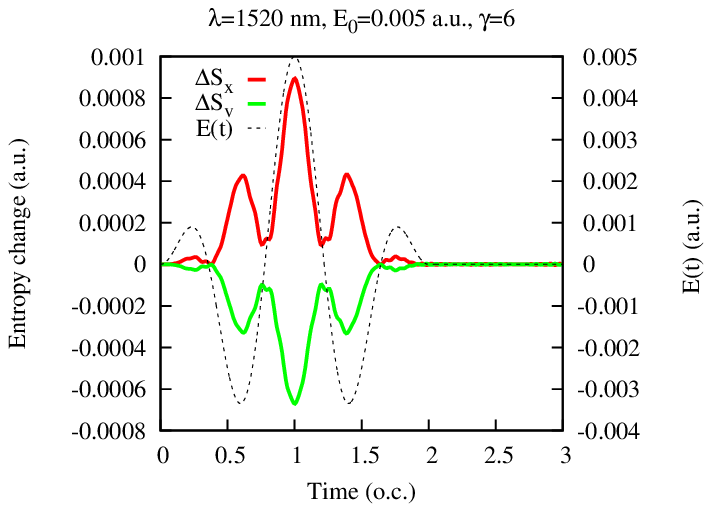}} 
\end{tabular}
\caption{Color online. Entropies of the coordinate and velocity distributions for a driving pulse
with base frequency $\omega=0.03$ a.u.  
for different electric field strengths. } 
\label{fig111}
\end{figure}

In \Fref{fig1} and \Fref{fig11}   we show the entropies of the coordinate and momentum distributions 
for different ionization regimes corresponding to different driving pulse frequencies.
We see
clearly distinct behavior of the entropies. With decreasing frequency the entropies 
begin to exhibit sharp variations at times close to the local maxima of the field, justifying the
expectations that entropies can capture the instant of time when  distributions begin to change. 
These variations 
are characteristic not only of
the tunneling regime of ionization.  They can appear also in the multiphoton regime  as \Fref{fig111} shows.
The difference  between the multiphoton regime in
\Fref{fig1} and \Fref{fig111}  is the 
large value of the  multiquantum parameter
$K=I_p/\omega$ in the case of the data shown in
\Fref{fig111}.  We cannot  expect a complete
similarity of the data for equal $\gamma$-values
if values of $K$ are different,   at least two
dimensionless parameters (e.g., $\gamma$ and
$K$ , or $\gamma$ and the reduced electric field $E_0/(2I_p)^{3/2}$)
are needed to  describe the ionization process
\cite{cpath3}.  The presence of the 
sharp variations of the entropies near the field maxima for large values of the $K$-parameter both in
\Fref{fig11} and \Fref{fig111} suggests that 
these variations are due to the high degree of 
non-linearity of the ionization process for the 
large $K$-values.

\Eref{ent} defining entropies is non-linear and rather
difficult to analyze. We will describe below some approximations,
which help to understand qualitatively the behavior of the 
entropies shown  in \Fref{fig1}, \Fref{fig11} and \Fref{fig111}.

Let us  represent the  time-dependent wave-function describing the 
evolution of the system  as:

\be
\displaystyle \Psi(t)=e^{-i\epsilon_0 t}\phi_0 + \Psi_1(t) \ ,
\label{p1}
\ee 

where $\phi_0$ is the initial ground atomic state with energy $\epsilon_0$. The \Eref{p1} serves as a definition of the 
wave-packet $\Psi_1(t)$.
It is  convenient to further decompose   $\Psi_1(t)$ 
in the mutually orthogonal components:

\begin{eqnarray}
\Psi_1(t)&=&\Psi_{\rm ion}(t)+\Psi_{\rm ex}(t)+\Psi_{\rm gs}(t)  \nonumber \\
&=&\Psi_2(t)+\Psi_{\rm gs}(t) ,\
\label{p2}
\end{eqnarray}

with $\Psi_2(t)= \Psi_{\rm ion}(t)+\Psi_{\rm ex}(t)$. 
These components are obtained by projecting  $\Psi_1(t)$
on the continuous spectrum of the field-free atomic Hamiltonian 
(in the case of $\Psi_{\rm ion}(t)$),  on the subspace spanned by  all excited bound states of  the hydrogen atom (for  $\Psi_{\rm ex}(t)$),  and on the 
subspace defined by the ground state of the  hydrogen atom (for $\Psi_{\rm gs}(t)$) . 

\begin{figure}[h]
\begin{tabular}{l}
\resizebox{100mm}{!}{\epsffile{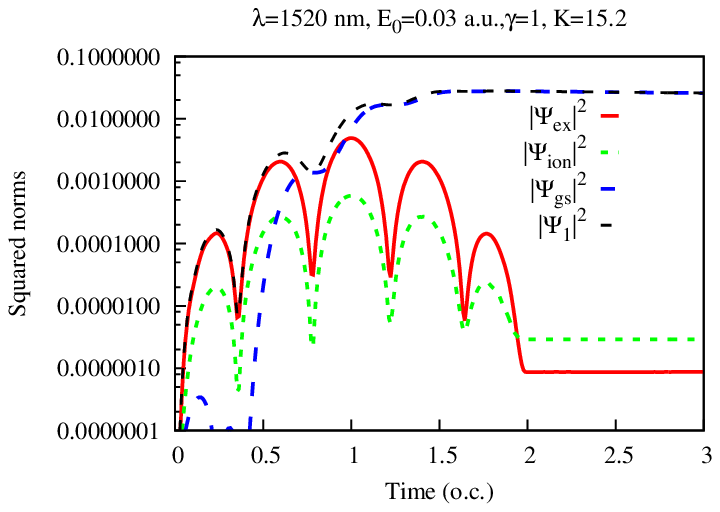}}  \\
\resizebox{100mm}{!}{\epsffile{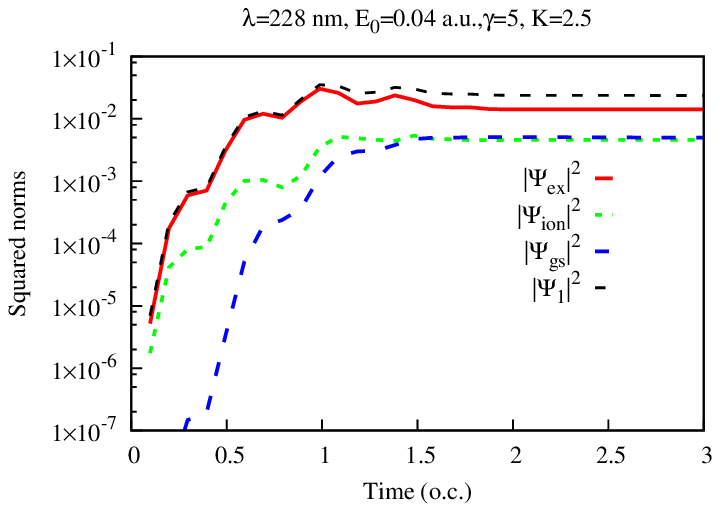}}  \\
\resizebox{100mm}{!}{\epsffile{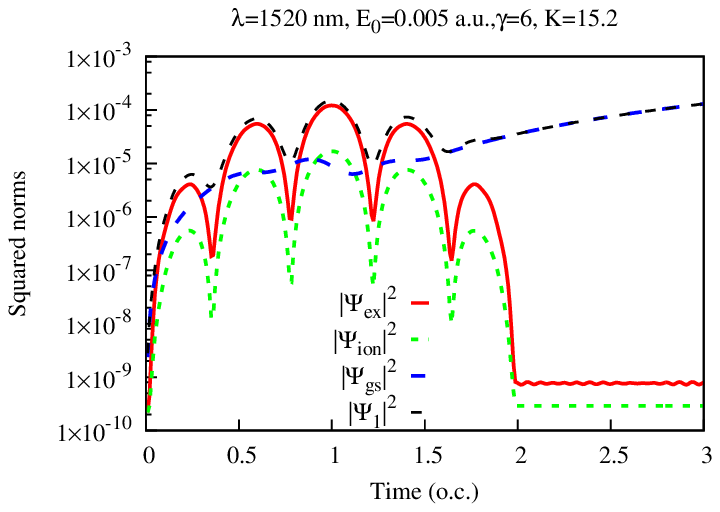}}
\end{tabular}
\caption{Color online.  Norms of the various components of
the wave-packet $\Psi_1(t)$  as 
functions of time  for different $\gamma$ and $K$ values. } 
\label{figex}
\end{figure}

Our goal is to use the  the decomposition \eref{p1}
of the total wave-function as a starting point for the perturbation expansion for the entropies.  For this perturbative approach
to work  we have to show 
that $||\Psi_1(t)||^2\ll 1$ in the course of the evolution.
To quantify the statement 
$||\Psi_1(t)||^2\ll 1$ in detail we show in \Fref{figex}
the squared norm of the wave-packet $\Psi_1(t)$
for the typical field parameters we employ in the calculations for different values of the 
multiquantum parameter $K$ and Keldysh parameter 
$\gamma$.  
For the reader's convenience we
show also the norms of the mutually orthogonal components
into which  $\Psi_1$ has been  decomposed in \Eref{p2}.
To avoid confusion, we note that only the norms
of the $|\Psi_{\rm ion}(t)|^2$ and 
$|\Psi_{\rm ex}(t)|^2$ remain constant after the 
end of the pulse, the norm  of
 the $\Psi_1(t)$, as defined
in the \Eref{p1}, generally varies even for the field-free
evolution.
As \Fref{figex} shows the norm
of $\Psi_1(t)$ satisfies  $||\Psi_1(t)||^2 \lesssim 0.025$ for 
all the interval of the pulse duration, 
thus justifying the applicability of the perturbative approach. 
As to the physical meaning of  norms of  the  components
into which  we decomposed $\Psi_1(t)$ above,  we
should emphasize, that 
this decomposition acquires  its full physical meaning 
only after the end of the laser pulse.  The squared norms $|\Psi_{\rm ion}(t)|^2$ and 
$|\Psi_{\rm ex}(t)|^2$ become, after the end of the pulse, 
the ionization 
and excitation probabilities, respectively. 
 Inside the interval of the pulse duration,   we cannot assign this
physical  meaning to $|\Psi_{\rm ion}(t)|^2$ and $|\Psi_{\rm ex}(t)|^2$  
quite unambiguously. This  can
be seen, e.g.,  from the fact that the separation \eref{p2} 
of the wave-function for the times inside the interval of the laser pulse
duration is not  gauge invariant.   It might be more correct
to talk not about the excitation and ionization 
processes for the times inside the laser pulse, but about 
the norms of the wave-packets 
$|\Psi_{\rm ion}(t)|^2$ and 
$|\Psi_{\rm ex}(t)|^2$. We will, however, use these terms interchangeably, even for the times inside the laser pulse
duration, which, we hope, will not lead to a confusion.
As one can
see from the \Fref{figex} and the corresponding 
plot in the \Fref{fig1}, \Fref{fig11},  the 
coordinate entropy largely mimics the 
behavior of the norms of the excitation and 
ionization probabilities. The reason for that is 
the highly nonlinear character of the ionization 
process for the large values of $K$.
We should emphasize, however,  
that unlike the notions
of the excitation and 
ionization for the times inside the interval of the 
laser pulse duration, the entropies are 
perfectly well-defined  quantities
which, therefore, can be used to follow 
the ionization process without any
ambiguities.

At the moment we are  more concerned
 with the mathematical  aspect of the problem. The statement 
$||\Psi_1(t)||^2\ll 1$ remains true for the weak enough fields we consider regardless of the choice of the gauge we may employ
to describe atom-field interaction. We will use this
fact to develop a perturbative approach to the
calculation of entropies, which will allow us to get
a more clear understanding of the entropies
behavior.  A detailed study of all the ionization regimes,
for different $\gamma$ and $K$ values being
hardly possible in a single publication, 
 we will concentrate below on the  case of the 
tunneling ionization for the field parameters 
used to obtain the data in the top panel of the \Fref{figex} ($\omega=0.03$ a.u., 
$E_0=0.03$ a.u.).

Let us consider the 
entropies changes $\displaystyle \Delta S_x(t)= S_x(t)-S_x(0)$ and
$\displaystyle \Delta S_v(t)= S_v(t)-S_v(0)$ induced by the field. Here, 
$S_x(0)$ and $S_v(0)$ are coordinate and velocity entropies of the field-free hydrogen atom in the 
ground state (these values 	can easily be calculated and are $S_x(0)= 3+ \log{\pi}\approx 4.15$,
$S_v(0)\approx 2.42$).
Expanding the integrand in \Eref{ent} using
$\displaystyle u\log{u} \approx
u_0\log{u_0}+ (u-u_0) (1+\log{u_0})$, with 
$u=|\Psi(t)|^2$,  $u_0=|\phi_0|^2$, and 
$u-u_0= |\Psi(t)|^2- |\phi_0|^2 \approx 2{\rm Re}
\ e^{i\epsilon_0 t} \phi^*_0 \Psi_1(t)$
and 
keeping only the terms up to the first order in the small 
$\Psi_1(t)$,  we
obtain:

\begin{eqnarray}
\Delta S_x(t) =  -2{\rm Re}\ e^{i\epsilon_0 t}\int  \phi^*_0(\r) \Psi_1(\r,t) \log{\left(e|\phi_0(\r)|^2\right)}\ d\r +
O(||\Psi_1(\r,t)||^2 ,\nonumber \\ 
\Delta S_v(t) =  -2{\rm Re}\ e^{i\epsilon_0 t}\int 
\tilde\phi^*_0({\bm q}) \tilde\Psi_1({\bm q},t) \log{\left(e|\tilde\phi_0({\bm q}|^2\right)}\ d{\bm q} 
+O(||\tilde\Psi_1({\bm q},t)||^2 .
\label{ent1}
\end{eqnarray}

\begin{figure}[h]
\begin{tabular}{l}
\resizebox{100mm}{!}{\epsffile{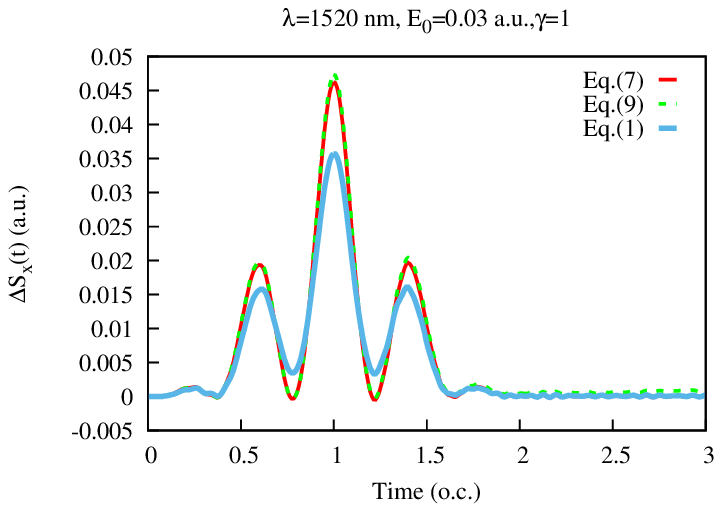}}  \\
\resizebox{100mm}{!}{\epsffile{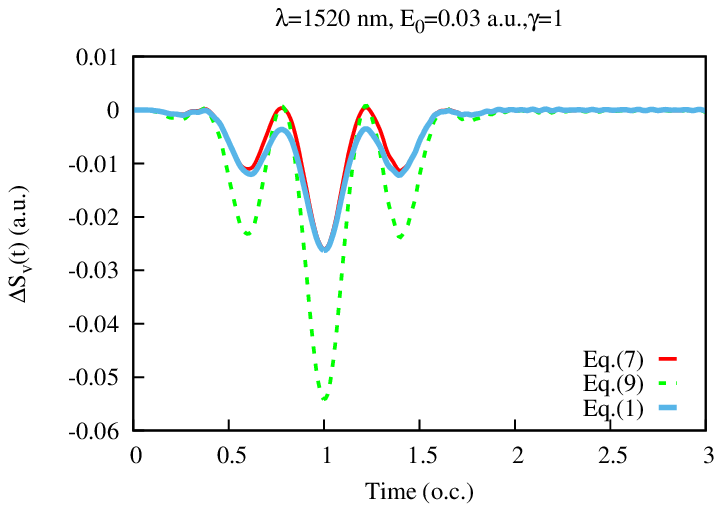}}  
\end{tabular}
\caption{Color online.  Results of the linearized (\Eref{ent1})
and asymptotic (\Eref{ent2}) approximations.}
\label{figlin}
\end{figure}

The tilted quantities in \Eref{ent1} stand for the Fourier transforms of the corresponding coordinate functions.  The accuracy of the 
linearized approximation \eref{ent1} can be judged from the 
\Fref{figlin}.

\begin{figure}[h]
\begin{tabular}{l}
\resizebox{80mm}{!}{\epsffile{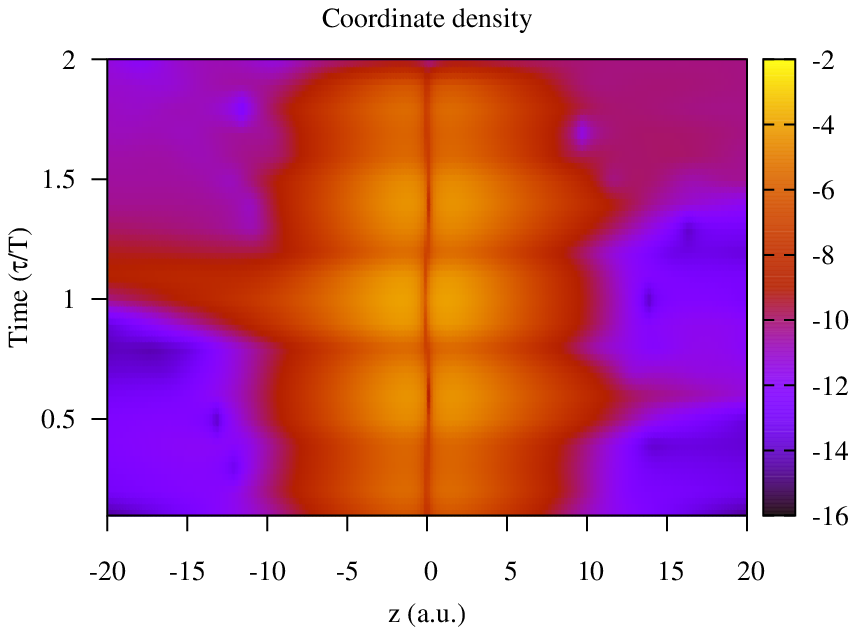}} \\
\resizebox{80mm}{!}{\epsffile{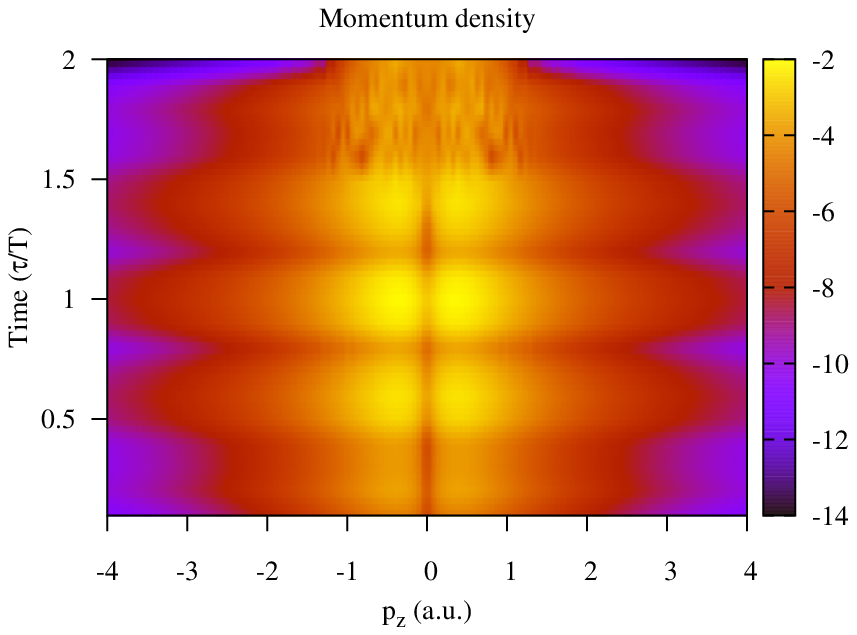}} 
\end{tabular}
\caption{Color online.  The coordinate and the momentum densities  (shown are the quantities  $\log_{10}{ |\Psi_2(0,0,z,t)|^2}$
and  $\log_{10}{|\tilde\Psi_2(0,0,p_z,t)|^2}$ as functions of time for the pulse with $E_0=0.03$ a.u., $\omega=0.03$ a.u.}
\label{fig22}
\end{figure}

To progress further, we note that 
the logarithmic factors in \Eref{ent1} vary slowly in the coordinate and momentum spaces.   Wave-packet  $\Psi_2(\r,t) $ defined
in \Eref{p2} and its Fourier transform 
$\tilde\Psi_2({\bm q},t) $,  on the other hand, vary fast in the 
coordinate and momentum spaces, correspondingly.  
To illustrate this statement,  we show in \Fref{fig22}  the
evolution in time of the coordinate density $|\Psi_2(0,0,z,t)|^2$,
and the momentum density  $|\tilde\Psi_2(0,0,p_z,t)|^2$ 
along the direction of the polarization vector. 
To show more structure,
which would be obscured had we used the linear
scale,  we use the  logarithmic scale in  \Fref{fig22}.

It is worthwhile to compare the behavior of the
coordinate density in \Fref{fig22} and the  norms of the wave-packets
$|\Psi_{\rm ion}(t)|^2$ and $|\Psi_{\rm ex}(t)|^2$  shown in the
\Fref{figex} which we discussed above.  The coordinate and the
momentum densities  in \Fref{fig22}
vary in time and reach maximum value every half cycle following the similar pattern in \Fref{figex}.   Let us take the interval 
around $t=1$ o.c. (the main maximum of the electric field) as
an example.  Horizontal slice of the top panel of  the \Fref{fig22}, taken
at $t=1$ o.c., shows a structure consisting of several components.
We can understand  these
structures qualitatively using the resolution   $\Psi_2(t)= \Psi_{\rm ion}(t)+ \Psi_{\rm ex}(t)$  of the wave-packet $\Psi_2(t)$ into the excitation and ionization parts we introduced in \Eref{p2}
We can interpret  the 
maxima of this structure located at $|z| \approx 1-2$ a.u. 
as the peaks due to the excitation part $\Psi_{\rm ex}(t)$ of the 
wave-packet.  The sleeve beginning at  $z \approx -10 $ a.u.  
can be attribute to the  "ionized" part $\Psi_{\rm ion}(t)$ of the 
wave-packet (for the laser pulse we consider electrons
can tunnel out in the negative $z$-direction at times close
to $t=1$ o.c. These interpretations are, of course, of 
a qualitative character only.   As we mentioned above, the 
different terms in the decomposition \Eref{p2} of the wave-packet  $\Psi_1(t)$  into its excitation ,
ionization, and the ground state  parts,  acquire the precise
physical meaning  only after the  end of the pulse.

The momentum density plot shown
in the bottom panel of \Fref{fig22} 
shows maxima which can again be attributed to the 
excitation  part of the wave-packet in the momentum 
space. One can also discern a ragged structure, which 
becomes dominant after $t \gtrsim 1.5$ o.c.  and which defines
the momentum distribution after the end of the pulse.
This ragged structure is due to the  ionized wave-packet.  
Such structures are known
to arise  in the longitudinal momentum distributions because of the
quantum interference \cite{arb3}. 
  
 Our aim in showing the densities in \Fref{fig22} was to 
 show that the wave-packets 
  $\Psi_2(\r,t) $ and $\tilde\Psi_2({\bm q},t) $ 
vary fast comparing to the logarithmic factors in the \Eref{ent1}.
Both coordinate and momentum distributions in \Fref{fig22} indeed change considerably  on a scale of an  atomic unit, 
both in the momentum and coordinate spaces).  The same observation applies
to the part $\Psi_{\rm gs}(t) $ of the wave-packet in \Eref{p2}
The well -known expressions for the coordinate and momentum space wave-functions of the hydrogen atom
$\displaystyle \phi_0(\r)={1\over\sqrt{\pi}}e^{-r}$, 
$\displaystyle \tilde \phi_0({\bm q})={\sqrt{8}\over \pi (1+q^2)^2}$
 \cite{LL3}  certainly  vary more slowly in the coordinate and momentum spaces  than their logarithms. 
 
For the 
estimate of the integrals in \Eref{ent1} we can use, therefore,
the well-known recipe \cite{fedoryuk},  often employed for the 
 asymptotic analysis of integrals. We can replace 
 the slowly varying logarithmic factors 
in \Eref{ent1} with their small -$r$ and small -$p$ asymptotics respectively.  These
asymptotics are: 
$\log{\left(e|\phi_0(\r)|^2\right)}  \approx   c_1 -2 r$,  $ \log{\left(e|\tilde\phi_0({\bm q}|^2\right)} \approx c_2 - 4q^2$, where $c_1=1$, $c_2=1+\log{8}-2\log{\pi}$.

Substituting these small -$r$ and small-$p$
expansions in the \Eref{ent1}  we obtain:

\begin{eqnarray}
\Delta S_x(t) =  -2{\rm Re}\ e^{i\epsilon_0 t}\int  \phi^*_0(\r) 
\Psi_1(\r,t) (c_1- 2r)\ d\r 
+O(||\Psi_1(\r,t)||^2 \  ,\nonumber \\ 
\Delta S_v(t) = -2{\rm Re}\ e^{i\epsilon_0 t}\int 
\tilde\phi^*_0({\bm q}) 
\tilde\Psi_1({\bm q},t)
(c_2-4q^2)\ d{\bm q}  
+O(||\tilde\Psi_1({\bm q},t)||^2  
\label{ent22}
\end{eqnarray}

The integral  $\int  \phi^*_0(\r)  r \Psi_{1}(\r,t)\ d\r$
in the \eref{ent22} for the coordinate entropy change 
is of the order of $O||\Psi_1||$.
To estimate the  integral $a(t)=
\int  \phi^*_0(\r)  \Psi_{1}(\r,t)\ d\r$ in the \Eref{ent22} 
we substitute for $\Psi_{1}(\r,t)$   the decomposition \eref{p2}:
$\Psi_{1}(\r,t)=\Psi_{2}(\r,t)+ \Psi_{\rm gs}(\r,t)$.
The integral  $\int  \phi^*_0(\r)  \Psi_{2}(\r,t)\ d\r$ 
is zero by the definition of $\Psi_{2}(\r,t)\ d\r$ in
\Eref{p2} and  orthogonality of the atomic states.
The magnitude of the remaining integral 
$a(t)= \int  \phi^*_0(\r)  \Psi_{\rm gs}(\r,t)\ d\r$ can be estimated as 
follows.  From  \Eref{p1} and \Eref{p2} we obtain:
$\Psi(t)= \left(e^{-i\epsilon_0 t}+a(t)\right)\phi_0+\Psi_2(t) \ $.
Using the orthogonality of $\phi_0$ and $\Psi_2(t)$
and normalization of $\Psi(t)$, one obtains then:
$|e^{-i\epsilon_0 t}+a(t)|^2+||\Psi_2(t)||^2=1$, from which it
follows that $2{\rm Re}\left( e^{i\epsilon_0 t} a(t)\right)+||\Psi_2(t)||^2+|a(t)|^2=0$.  Since
$|a(t)|=O||\Psi_1(t)||$ it follows that
$2{\rm Re}\left( e^{i\epsilon_0 t} a(t)\right)=O||\Psi_1(t)||^2$.
This  means that the term containing 
$a(t)= \int  \phi^*_0(\r)  \Psi_{\rm gs}(\r,t)\ d\r$ in the 
\Eref{ent22}  is small compared to other terms in this equation and can be dropped. The same estimates apply in the case of the expression
for the change of the velocity entropy in \Eref{ent22}. We
arrive, thus, at the following approximate 
expressions for the entropy changes :

\begin{eqnarray}
\Delta S_x(t) =  4{\rm Re}\ e^{i\epsilon_0 t}\int  \phi^*_0(\r) \Psi_1(\r,t) r\ d\r 
+O(||\Psi_1(\r,t)||^2 \  ,\nonumber \\ 
\Delta S_v(t) = 8{\rm Re}\ e^{i\epsilon_0 t}\int 
\tilde\phi^*_0({\bm q}) \tilde\Psi_1({\bm q},t)q^2\ d{\bm q} 
+O(||\tilde\Psi_1({\bm q},t)||^2 \ .
\label{ent2}
\end{eqnarray}

The  approximations we have introduced so far to simplify
the exact equation \Eref{ent}  were the linearized perturbation expression 
\eref{ent1} and its further simplification \eref{ent2}, obtained 
by replacing logarithmic  factors with leading terms of their
Taylor expansions, which give non-zero contributions. 
The \Fref{figlin}  illustrates how these approximations 
actually work. One can see that linearized \Eref{ent1} represents
exact entropies pretty accurately on the whole time interval we
consider.  \Eref{ent2} represents variation of the coordinate 
entropy quite accurately, but is less accurate in the case of the 
velocity entropy. This is probably because
the faster growing factor $q^2$ in the \Eref{ent2} 
for the velocity entropy makes 
the approximation based on the assumption 
that only the small vicinity near the origin contributes to 
the integral less accurate.  In addition, as we saw above, 
$\tilde\Psi_2({\bm q},t) $ is highly oscillatory for 
$t \gtrsim 1.5$ o.c. 
We may expect the accuracy of an asymptotic estimate based
on using the leading term of the expansion of the logarithmic
factor in \Eref{ent1} near the origin to be less accurate for
the oscillatory $\Psi_2$.  Indeed,  
integrals of  rapidly 
oscillating functions are known to be more difficult to evaluate
accurately, because such integrals typically have 
small values,  even when integrands are relatively large.
\Fref{figlin} shows, however, that even in the case of the 
velocity entropy we can still rely on the  \Eref{ent2}  at least for 
qualitative estimates.

\begin{figure}[h]
\begin{tabular}{ll}
\resizebox{80mm}{!}{\epsffile{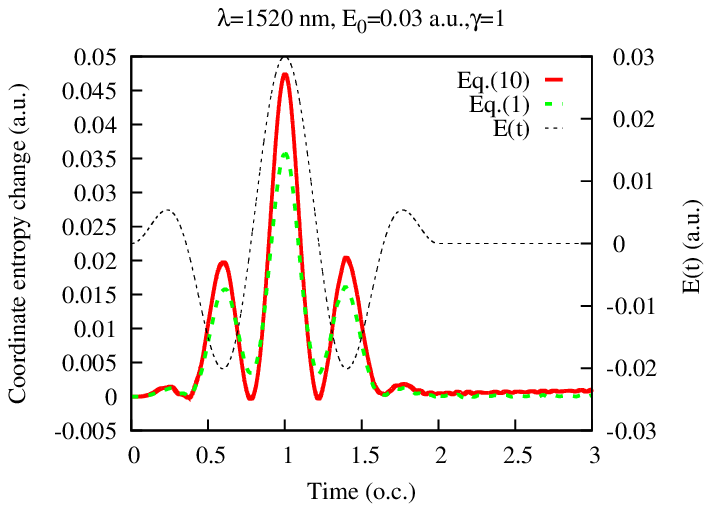}}   &
\resizebox{80mm}{!}{\epsffile{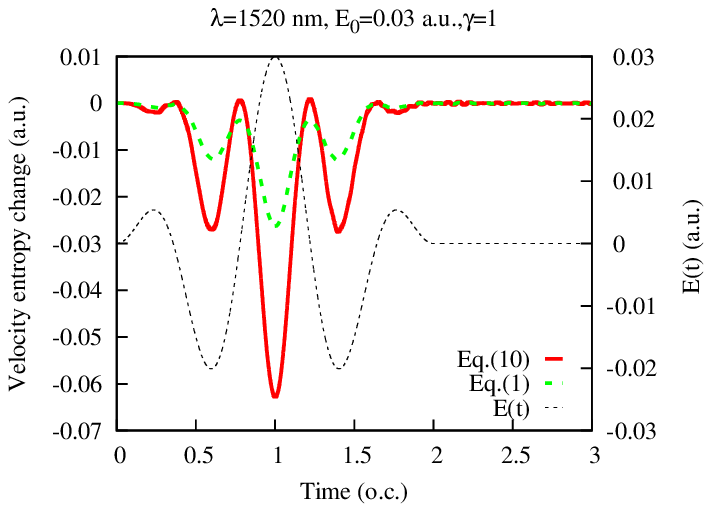}}  \\ 
\resizebox{80mm}{!}{\epsffile{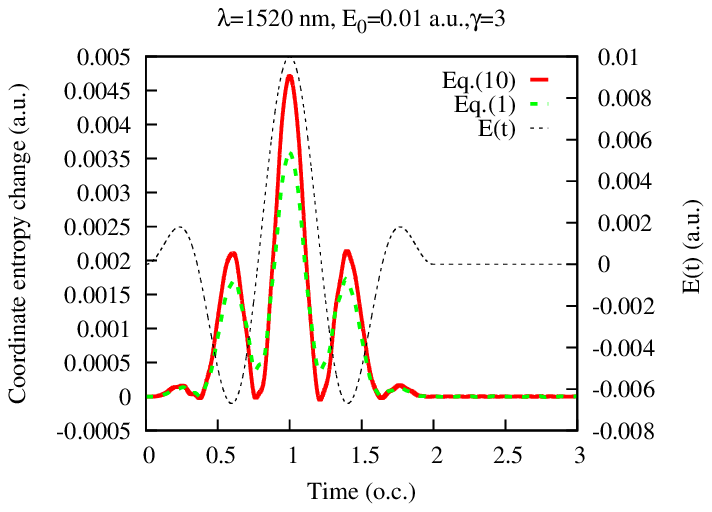}}   &
\resizebox{80mm}{!}{\epsffile{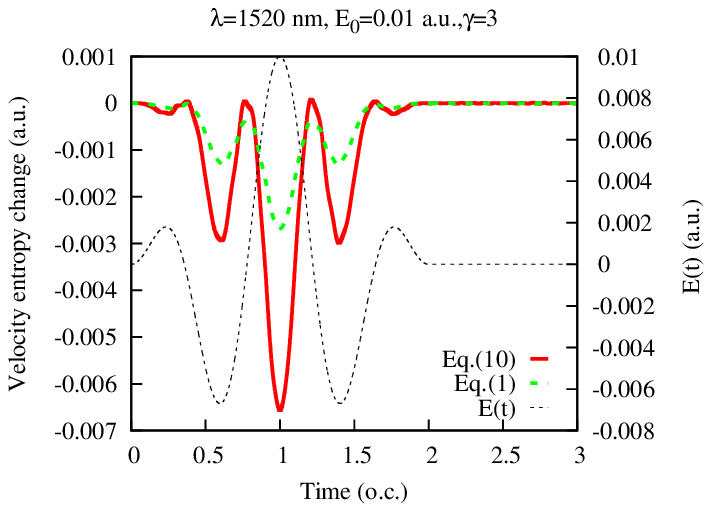}} 
\end{tabular}
\caption{Color online. Exact (\Eref{ent}  and approximate (\Eref{ent3}) 
entropies of the coordinate and velocity distributions.} 
\label{figkin}
\end{figure}

To conclude this discussion
 we note that  the \Eref{ent2} can
be approximately rewritten in yet another useful form as:

\begin{eqnarray}
\Delta S_x(t) =  2 \Delta\langle \Psi(\r,t)|r|\Psi(\r,t)\rangle  \ + 
O(||\Psi_1(\r,t)||^2 ,\nonumber \\ 
\Delta S_v(t) = 4 \Delta\langle\tilde \Psi({\bm q},t)|q^2|\tilde\Psi({\bm q},t)\rangle+O(||\tilde\Psi_1({\bm q},t)||^2 \ ,
\label{ent3}
\end{eqnarray}

where symbol $\Delta$ on the r.h.s. means the difference of the expectation value 
at time $t$ and the initial moment of time.
To show that this equation is approximately valid, 
one can use the decomposition \eref{p1}
and see, keeping the terms linear in $\Psi_1$,
that r.h.s of \Eref{ent3} and l.h.s of \Eref{ent2} coincide.  Results
given by the  \Eref{ent3} for the 
coordinate and velocity entropies are shown in  the \Fref{figkin}.

Being a direct consequence of the \Eref{ent2},  approximation formulas
\eref{ent3} naturally inherit their main features. As 
\Fref{figkin} shows \Eref{figkin},  provides a pretty accurate
approximation in the case of the coordinate entropy, and 
less accurate, but still qualitatively useful, approximation for the
velocity entropy.

To see why the coordinate and velocity entropies
attain, correspondingly, their local maxima
and minima at the times near the 
local field maxima in our picture, we can take a closer look
at the \Eref{ent3}. We introduced above the
partial wave expansion for the coordinate 
wave-function $\Psi(\r,t)$.   Using analogous 
expansion for the momentum space wave-function
 $\tilde \Psi({\bm q},t)$:
 
\be
\tilde \Psi({\bm q},t)=
\sum\limits_{l=0}^{l_{\rm max}} 
{g_{l}(q,t)\over q} Y_{l0}(\hat{\bm q}) \ ,
\label{basisq}
\ee

we can rewrite \Eref{ent3} in the following form:

\begin{eqnarray}
\Delta S_x(t) =  
2 \int\limits_0^{\infty} \left(|f_0(r,t)|^2-|f_0(r,0)|^2\right)r\ dr + O(||\Psi_1(\r,t)||^2 ,\nonumber \\ 
\Delta S_v(t) = 
4 \int\limits_0^{\infty} \left(|g_0(r,t)|^2-|g_0(r,0)|^2\right)q^2\ dq  +O(||\tilde\Psi_1({\bm q},t)||^2 \ ,
\label{ent33}
\end{eqnarray}

\begin{figure}[h]
\begin{tabular}{ll}
\resizebox{80mm}{!}{\epsffile{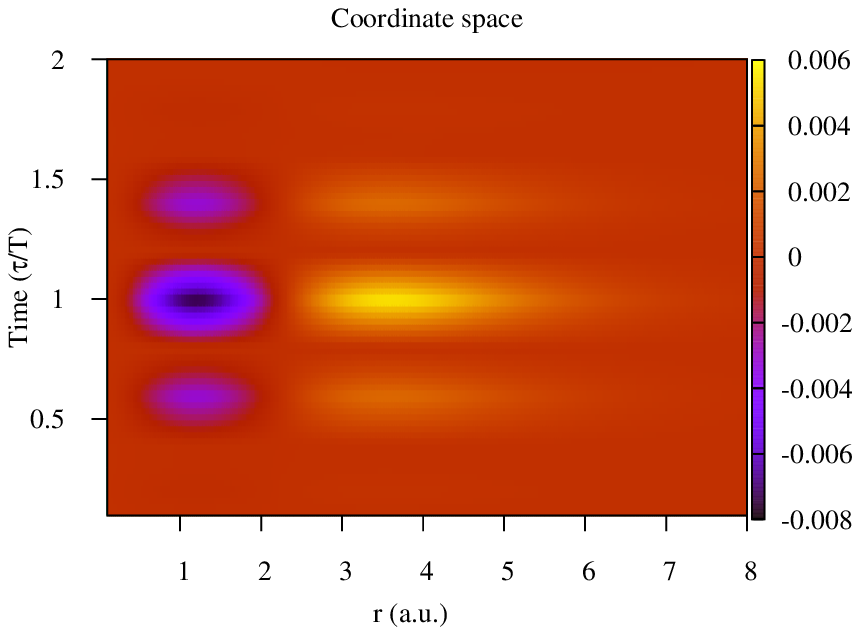}}  &
\resizebox{80mm}{!}{\epsffile{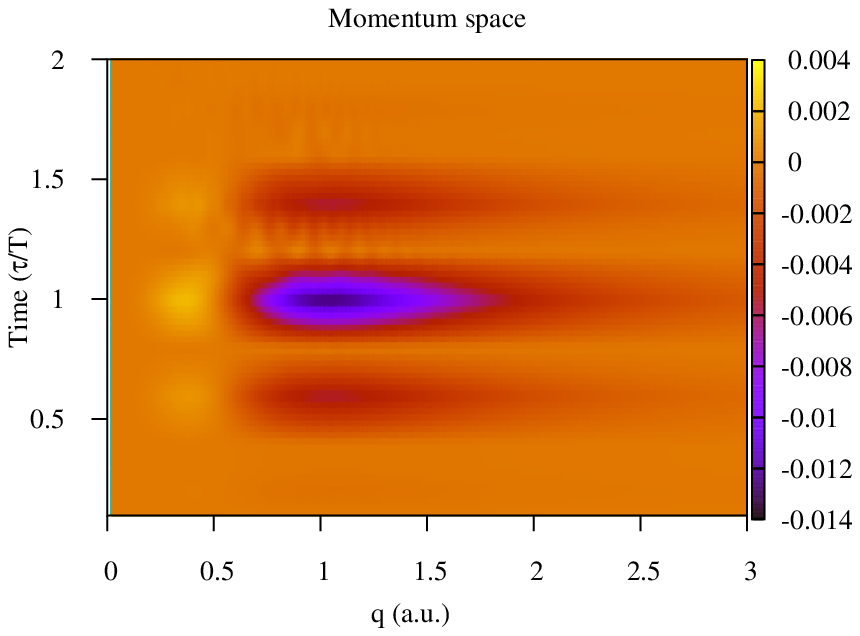}} \\
\resizebox{75mm}{!}{\epsffile{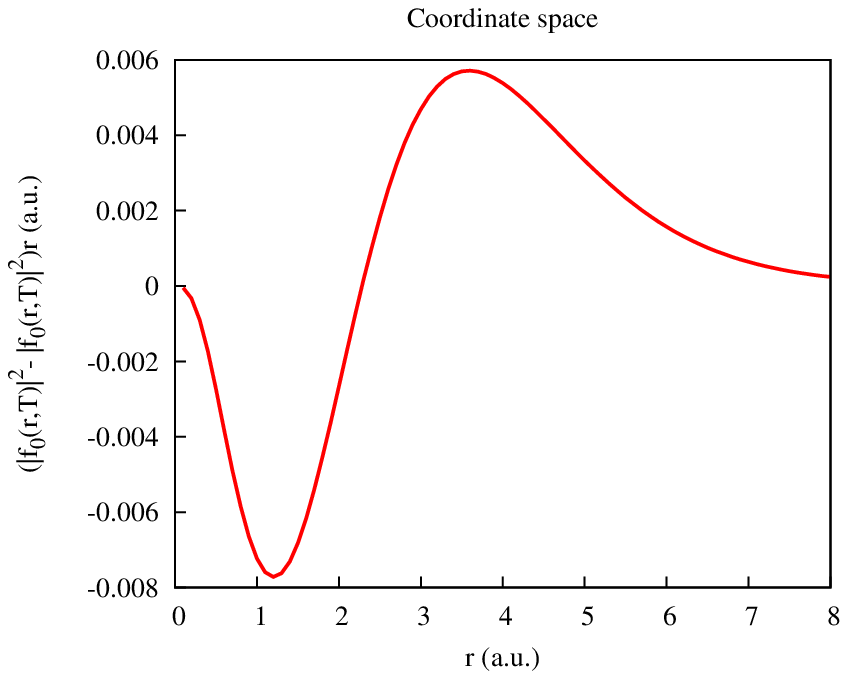}} &
\resizebox{75mm}{!}{\epsffile{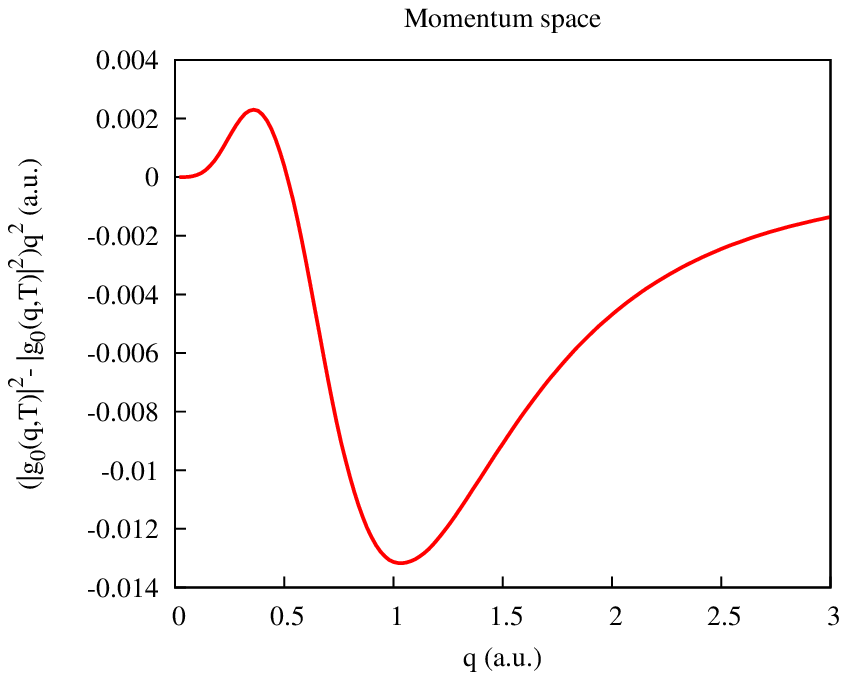}}
\end{tabular}
\caption{Color online.  Integrands in 
\Eref{ent33} in  the coordinate and momentum
spaces as functions of time (top row) and 
for $t=1$ o.c. (bottom row)
for the pulse with $E_0=0.03$ a.u.,
 $\omega=0.03$ a.u .}
\label{figd0}
\end{figure}

Evolution of the integrands in the  \Eref{ent33} in time is shown in 
\Fref{figd0}.  
The processes which influence the wave-function of the 
system are the excitation and ionization processes. Both these 
processes lead to the flow of the probability density.  This flow,
however manifests itself differently in the coordinate
and momentum spaces. 
Let us consider first the case of the momentum space,
and the velocity entropy for  the times near the
main maximum of the 
field ($t=1$ o.c).  As one can see from  the \Fref{figd0}, the 
 integrand in 
the corresponding equation \eref{ent33} has  a small positive
maximum  at $q\approx 0.5$ a.u., and a much deeper minimum 
at $q\approx 1$.  As the horizontal slice 
of the momentum distribution in \Fref{figd0} shows, it
it this structure which makes the corresponding integral in 
\Eref{ent33}, and hence the velocity entropy change at
$t\approx 1$ o.c.,  negative. 
The origin of this structure can be explained 
qualitatively as 
follows.  In the momentum  space both ionization and 
excitation lead to the flow of the probability from larger to 
 smaller $q$-values. This happens because in the momentum 
 space both ionization and 
 excitation processes produce  components in the wave-function which
 are more sharply peaked at the origin than the initial state  wave-function.
For the excited states it follows from the trivial fact that  the 
more a  function gets extended in the coordinate space, 
the more narrow its Fourier transform becomes. 
As far as the ionization process is concerned , it  
produces, according to the SFA \cite{Keldysh64,tunr,tunr2},
a momentum distribution of the ionized
electrons which is a Gaussian localized near the origin in the momentum space. This Gaussian again is a more narrow distribution than the 
momentum distribution in the ground state of hydrogen. 
Both these processes, therefore, lead to the effective probability
flow to the region of the  small $q$-values, thus making the 
corresponding integral in the  \Eref{ent33}, and the change
of the velocity entropy, negative in the vicinity of 
$t\approx 1$ o.c..

In the coordinate space the same processes lead to a 
different picture.  The excitation process, which as we have seen above in the \Fref{figex}, is the dominant process for the times inside 
the interval of the laser pulse
duration,  produces components in the wave-function which 
are more extended in the coordinate space than the 
initial state.  As far as the ionization process is concerned,
the part of the  wave-packet in the \Eref{p2}  describing 
ionized electron is localized  in the coordinate space 
near the "exit point". As an estimate for the coordinate
of the exit point we can use either 
the adiabatic tunnel exit point value $|z_e| \approx {I_p\over E_0}$,
or the Field Direction Model (FDM) expression \cite{landsman2015} $|z_e| \approx {I_p +\sqrt{I_p^2-2E_0}\over E_0}$, which give 
$|z_e| \approx 16$ a.u.  for the field strength of $0.03$ a.u.
Alternatively, we can  estimate the exit point as the sleeve position 
in \Fref{fig22}, showing coordinate distribution for the same 
field strength, which gives $|z_e|\approx 10$ a.u.
Employing either estimate we obtain an additional component in
the wave-function increasing coordinate density at
 large distances from the origin. The probability in the 
 coordinate space, therefore, flows in the opposite 
 direction,  from the small to the large distances. That makes the 
corresponding integral in \Eref{ent33}, and hence the coordinate
entropy change, positive.

Other  features of the behavior of the entropies in \Fref{fig1}, 
\Fref{fig11}, and \Fref{fig111}
we have to explain,  are  the decrease of the coordinate entropy
and increase of the velocity entropy for the times after the
ionization event. 
As we have seen in \Fref{figex}, the excitation and ionization
processes, when we consider them for the times inside the laser
pulse duration, are not irreversible.  After we pass the 
field maximum and field strength starts to decrease, the  excitation and ionization probabilities
diminish.  In the picture we developed above that means that
the probability  flows in the coordinate and momentum spaces
reverse their directions. The coordinate entropy, therefore,
 decreases while the  velocity entropy increases. 
 
 These arguments can be also used to account for the  visible 
(\Fref{fig1},\Fref{fig11})  net gain of the coordinate, and loss of the velocity entropies towards  the end of the pulse, compared to their values
in the initial ground atomic state. The excitation and ionization
probabilities  tend toward the 
end of the pulse to their physical values.   As shown in \Fref{figex},
these physical values are much less than the values 
of the corresponding squared norms  
$\Psi_{\rm ex}(t)$, $\Psi_{\rm ion}(t)$  for the times inside the interval
of the laser pulse duration.  Their effect, however, is the same as
we described above, excitation and ionization lead to the 
changes of the probability densities in the coordinate and momentum
spaces which by the mechanism encapsulated in \Eref{ent33}
lead to the increase of the coordinate and decrease of the velocity
entropies. The magnitude of the effect at the end of the pulse  is, however,  much smaller, hence the smaller net increase of the coordinate entropy, and smaller decrease of the velocity entropy toward the end of the pulse, compared to the magnitude of their changes  for 
the times inside the interval of the pulse duration.

\begin{figure}[h]
\begin{tabular}{ll}
\resizebox{80mm}{!}{\epsffile{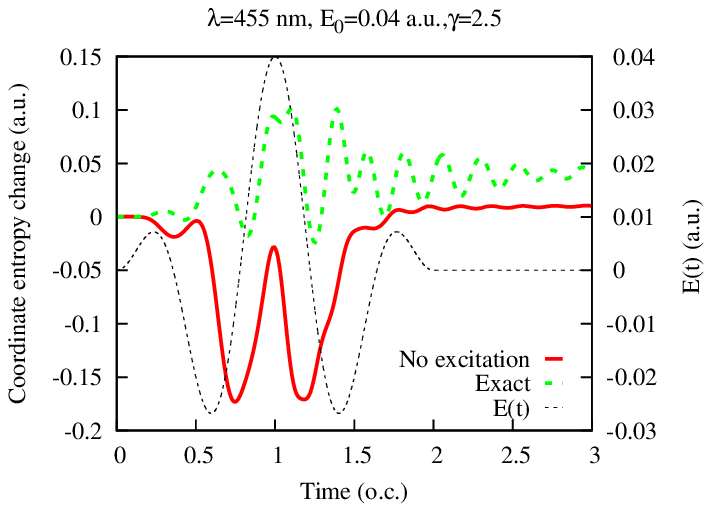}}   &
\resizebox{80mm}{!}{\epsffile{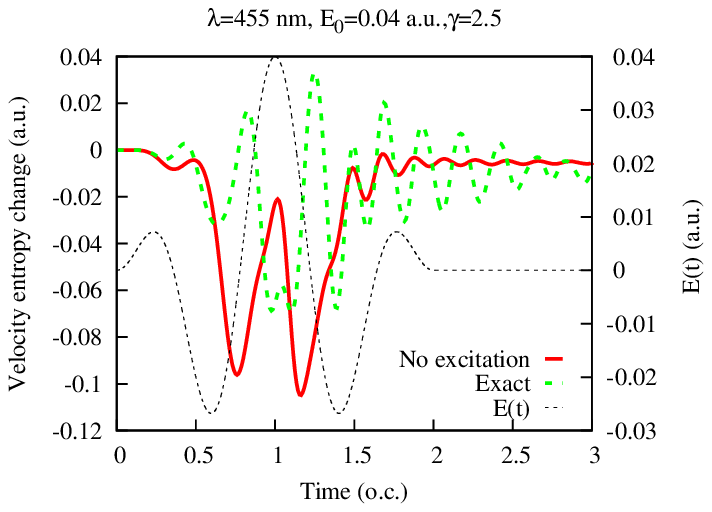}}  \\ 
\resizebox{80mm}{!}{\epsffile{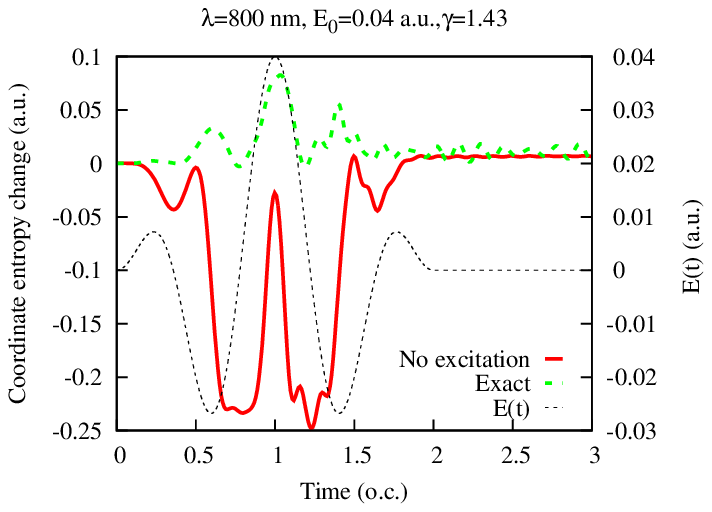}}   &
\resizebox{80mm}{!}{\epsffile{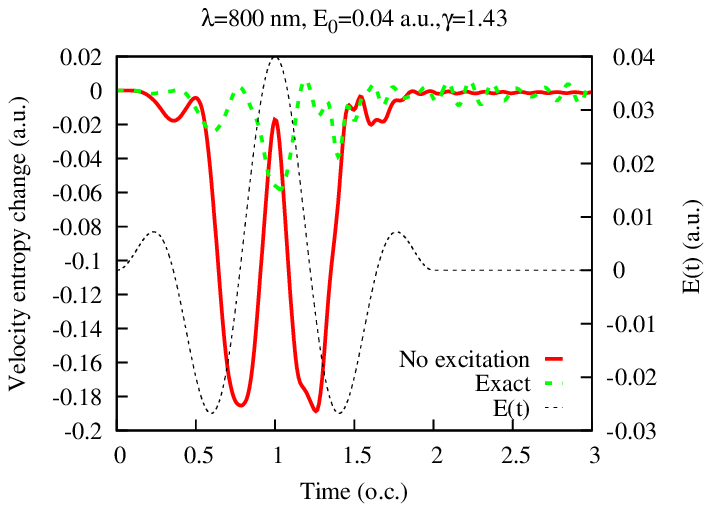}}  \\ 
\end{tabular}
\caption{Color online. coordinate and velocity entropies 
for the calculations using  the complete TDSE  wave-function and 
the wave-function with the excitation channel projected out.} 
\label{fignex}
\end{figure}

The features of the entropies in
 \Fref{fig1}, 
\Fref{fig11}, and \Fref{fig111},   we must explain yet, are the oscillations 
clearly visible after the end of the pulse.  
These oscillations are due to the population 
trapped in the excited states after the end of the pulse. One can see from \Eref{ent1} that a component in  
$\Psi_1(\r,t)$  or  $\tilde \Psi_1({\bm q},t)$
due to an excited state with 
energy $\epsilon_1$ leads to the terms in  the entropies
oscillating with time with frequency $\epsilon_1-\epsilon_0$,
provided the excited state component in 
$\Psi_1$ has $s$-symmetry (otherwise the integral in 
\Eref{ent1} would be zero for the initial $s$-state).  Amplitude of the 
oscillations is defined by the norm of the corresponding component
of $\Psi_1$. 
To further elucidate this issue, we show in the \Fref{figex}
the coordinate and velocity entropies obtained if
we project out of the wave-function $\Psi(t)$ the component
due to the excited states. In other words, to calculate entropies
we use \Eref{ent}  not with the full TDSE wave-function 
$\Psi(t)$,  but with the wave-function $(\hat I-\hat Q_{\rm ex})\Psi(t)$,
where $\hat Q_{\rm ex}$ is the projection operator on the 
excited bound states of the hydrogen atom.  As 
\Fref{fignex} shows, using this procedure we indeed 
get rid of the oscillations in the entropies after the  end of the 
pulse. Behavior of the entropies inside the pulse changes too,
of course. As the \Fref{figex} shows the excitation channel is,
in fact, dominating for times inside the interval of the laser pulse duration.

Exact entropies in the \Fref{fignex}  oscillate with 
periods of approximately  $T/4$ ( the case shown in the top row of
\Fref{fignex}  and $T/6$ (bottom panel  of the Figure). 
Corresponding energy differences  $\epsilon_1-\epsilon_0$  are
$4 \omega \approx 0.4$ a.u.,  and $6 \omega\approx 0.35$ a.u., respectively ($T$ and $\omega$ here are the
optical cycle duration and the base frequency in each case).  
According to this estimate, the oscillations in both cases are due to the 
excitations of the $2s$ state of hydrogen, 
which is a four photon process
for $\omega=0.1$ a.u. , 
and a six photon process for  $\omega= 0.057$ a.u.   
The latter excitation process  is a much weaker one,  hence
the much smaller amplitude of oscillations for the
case of $\omega= 0.057$ a.u.  in \Fref{fignex}. 
Oscillations of the entropies after the end of the pulse can provide, 
therefore, information about the excitation mechanism.


\section{Conclusion}

To conclude, we have used the information theoretic entropies of coordinate and velocity distributions to 
follow  the evolution of the  wave-function 
during the strong field ionization process. 
The property of the information theoretic entropies we employed is their ability
to supply the space of probability distributions
with a metric and to gauge the "distance" between different probability distributions.
This property allows to visualize conveniently the evolution of the quantum mechanical distributions
and to locate the instants of time when distributions start to evolve appreciably. 

We considered
both multiphoton and tunneling regimes of ionization. 
We saw that with increasing multiquantum parameter
$K$ the   entropies 
begin to exhibit increasingly sharper variations at times close to the local maxima of the field.
Behavior  of the entropies  after the end of the laser pulse 
provides information about the excited states population and the excitation process.

As we mentioned above, the important notion of the Keldysh tunneling time 
can be introduced as the time the wave-function of the system takes to adjust 
when the electric field is turned 
on instantly \cite{time_adjust}. In the present context, the non-zero tunneling time would result
in the lag between the instance when wave-function of the system starts to evolve, and the maxima of the 
electric field. As we have seen,  study of the behavior of the 
entropies  as functions of time allows to capture 
the instances of the rapid evolution of the wave-function rather precisely.  We can adopt a  definition of the tunneling time
as the lag between the local maximum of the electric field and
the extremum of the coordinate or velocity entropy.
As we saw above,  entropy increase near the 
peak of the electric field is due to both excitation and
ionization processes (as we noted, this
division has only a qualitative character inside the
laser pulse).   As the top panel of the \Fref{figex} shows, near the main maximum of the 
electric field the contribution of the 
ionization  channel is of the order of about 10
percent of the contribution of the excitation 
channel. This amount would be  enough to produce an appreciable lag in the entropies, if any lag in the 
ionization channel were present. This definition 
of the tunneling delay, as the lag between the
contribution of the ionization channel and
the electric field was adopted in \cite{inst1}.
The contribution of the ionization channel was
found using the same procedure we employed  
in our work, 
by projecting out contributions of the bound states from the solution of the TDSE. The difference with the
present approach is the gauge-dependent character
of this procedure.  Entropies, on the other hand
are gauge independent.  
Our calculations show no  lag of any appreciable
value for the entropies
and hence zero tunneling delay in the framework of the definition we proposed above. We find, for instance, that for the  pulse parameters in the top panel of the \Fref{figex}: $\omega=0.03$ a.u., $E_0=0.03 $ a.u., the coordinate
entropy lags behind the main maximum of the 
electric field by approximately $0.0016$ o.c., and the 
velocity entropy by approximately $0.0011$ o.c.
These numbers  are well within the numerical uncertainty of our calculations
and lend support to some other results in the 
literature claiming essentially zero tunneling delay 
 \cite{tor,Nie,Nie1}. 
 
Finally, we would like to emphasize the utility of the 
the entropy based view.  It is true, of course, that all the
 information about the ionization process is, in principle, available in the wave-function.  It may be not easy, however, to extract and
 analyze this  information.  The wave-function (or various densities which can be derived from it) are , in general,  functions of three variables and time. Evolution of such an object is not easy to follow.  The entropy, on the other hand,  is a single number which,  as we hope we were able to demonstrate,  encapsulates the key changes  the system 
can undergo during the ionization process.

\section{Acknowledgments} 

This work was supported by the Institute for Basic Science
under the grant number IBS-R012-D1.

\newpage


\end{document}